\documentclass[aps,prb,showpacs,floatfix,twocolumn,byrevtex]{revtex4-1}
\pdfoutput=1
\usepackage{amstext}
\usepackage{amsmath}
\usepackage{amsfonts}
\usepackage{color}
\usepackage{dcolumn}
\usepackage{graphicx}
\usepackage{hyperref}
\usepackage{bm}
\begin{document}

%\preprint{Draft, not for distribution}
%
% The title and the list of authors
%
%\title{Hint of a condensate in the optical properties of K$_{\mathbf{0.8}}$Fe$_{\mathbf{2-y}}$Se$_{\mathbf{2}}$}
\title{Optical conductivity of superconducting K$_{\mathbf{0.8}}$Fe$_{\mathbf{2-y}}$Se$_{\mathbf{2}}$
 single crystals: \\ Evidence for a Josephson-coupled phase}
\author{C. C. Homes}
\email{homes@bnl.gov}
\author{Z. J. Xu}
\author{J. S. Wen}
\author{G. D. Gu}
\affiliation{Condensed Matter Physics and Materials Science Department,
  Brookhaven National Laboratory, Upton, New York 11973, USA}%
\date{\today}

%
% The abstract goes here
%
\begin{abstract}
The optical properties of the iron-chalcogenide superconductor K$_{0.8}$Fe$_{2-y}$Se$_2$ with
a critical temperature $T_c = 31$~K have been measured over a wide frequency range in the
{\em a-b} planes above and below $T_c$.  The conductivity is incoherent at room temperature,
but becomes coherent (Drude-like) at $T\gtrsim T_c$; however, $R_\Box \simeq 320$~k$\Omega$,
well above the threshold for the superconductor-insulator transition at $R_\Box = h/4e^2
\simeq 6.9$~k$\Omega$.  Below $T_c$, the superfluid density $\rho_{s0} \simeq 48 \times
10^3$~cm$^{-2}$ places this material on the scaling line $\rho_{s0}/8 \simeq 4.4\,
\sigma_{dc}\, T_c$, but in a region associated with Josephson coupling, suggesting this
material is inhomogeneous and constitutes a Josephson phase.
\end{abstract}
%
%  PACS numbers
%  63.20.-e  Phonons in crystal lattices
%  89.75.Da Systems obeying scaling laws
%  72.15.-v Electronic conduction in metals and alloys
%  72.80.-r Conductivity of specific materials
%
%  74.25.Bt SC: Thermodynamic properties
%  74.25.Gz SC: Optical properties
%  74.25.Ha SC: Magnetic properties
%  74.25.Nf SC: Response to electromagnetic fields
%  74.70.-b SC:	Superconducting materials other than cuprates
%  74.72.Bk SC: Y-based cuprates
%  74.81.-g SC:	Inhomogeneous superconductors and superconducting systems, including electronic inhomogeneities
%
%  77.22.Ch  Permittivity (dielectric function)
%  78.20.-e  Optical properties of bulk materials and thin films
%  78.30.-j  Infrared and Raman spectra
%
\maketitle
%
% The main body of the text
%
% Introduction
%
The discovery of superconductivity in the iron-arsenic (pnictide) material
LaFeAsO$_{1-x}$F$_x$\cite{kamihara08} and the rapid increase in the
critical temperature $T_c$ above 50~K through rare-earth substitutions,\cite{ren08}
has prompted the search for related iron-based superconductors in
the hope of achieving even higher values for $T_c$. While superconductivity
was quickly discovered in (Ba$_{1-x}$K$_x$)Fe$_2$As$_2$\cite{rotter08b} and
the iron-chalcogenide FeTe$_{1-x}$Se$_x$,\cite{fang08} the maximum value
for $T_c$ in these compounds is $\simeq 38$ and 15~K, respectively.
In the iron-based superconductors, scattering between hole and
electron pockets in this class of materials is considered a necessary
element for high critical temperatures.\cite{chubukov12}  Indeed,
in the KFe$_2$As$_2$ material, the electron pockets that
are present in all the other iron-based materials are absent,
leaving just the hole pockets at the center of the Brillouin
zone and a severely reduced $T_c \simeq 3$~K.\cite{chen09,sato09}
The discovery of superconductivity in K$_{0.8}$Fe$_{2-y}$Se$_2$ with
$T_c \gtrsim 30$~K\cite{guo10} was greeted with enthusiasm not only
because of the relatively high value for $T_c$, but also as a result
of its unique electronic structure.  In this material the hole
pockets at the center of the Brillouin zone are absent, leaving just
the electron pockets at the edges of the zone;\cite{x-wang11,zhang11,qian11}
however, the value for $T_c \simeq 30$~K is an order of magnitude
larger than in the hole-doped analog, suggesting that the spin
fluctuation pairing mechanism may have to be re-evaluated.\cite{f-wang11}
In the superconducting state K$_{0.8}$Fe$_{2-y}$Se$_2$ displays
a nearly isotropic gap of $8-10$~meV on the Fermi
surfaces.\cite{x-wang11,zhang11,qian11}
Evidence for phase separation and the coexistence of magnetism and
superconductivity has been observed in recent optical work on
K$_{0.75}$Fe$_{1.75}$Se$_2$\cite{yuan11} and the related
Rb$_2$Fe$_4$Se$_5$ material;\cite{charnukha12} this interpretation
is consistent with other recent experiments.\cite{z-wang11,ricci11a,ricci11b,shermadini11}

%
% What is new in this work...
%
In this Rapid Communication we present the detailed in-plane optical properties of single
crystal of K$_{0.8}$Fe$_{2-y}$Se$_2$ ($T_c = 31$~K) in both the normal and
superconducting state.  At room temperature the optical properties are
dominated by the infrared-active vibrations and other bound excitations;
the free-carrier response is incoherent.  As the temperature is reduced
a coherent, or Drude-like, response is observed just above $T_c$.  Below
$T_c$ a weak superconducting response is observed allowing the superfluid
density $\rho_{s0}$ to be estimated.  This material is observed to fall on
the general scaling line for the cuprate superconductors, $\rho_{s0}/8
\simeq 4.4 \sigma_{dc} T_c$,\cite{homes04} but in a region typically
associated with the {\em c}-axis response where the superconducting
response is due to Josephson coupling, suggesting that this material
is indeed phase separated and constitutes a Josephson phase.\cite{imry08}

%
% Figure 1 - Reflectance and unit cell.
%
\begin{figure}[b]
%\centerline{\includegraphics[width=2.8in]{figure1.eps}}%
\includegraphics[width=0.90\columnwidth]{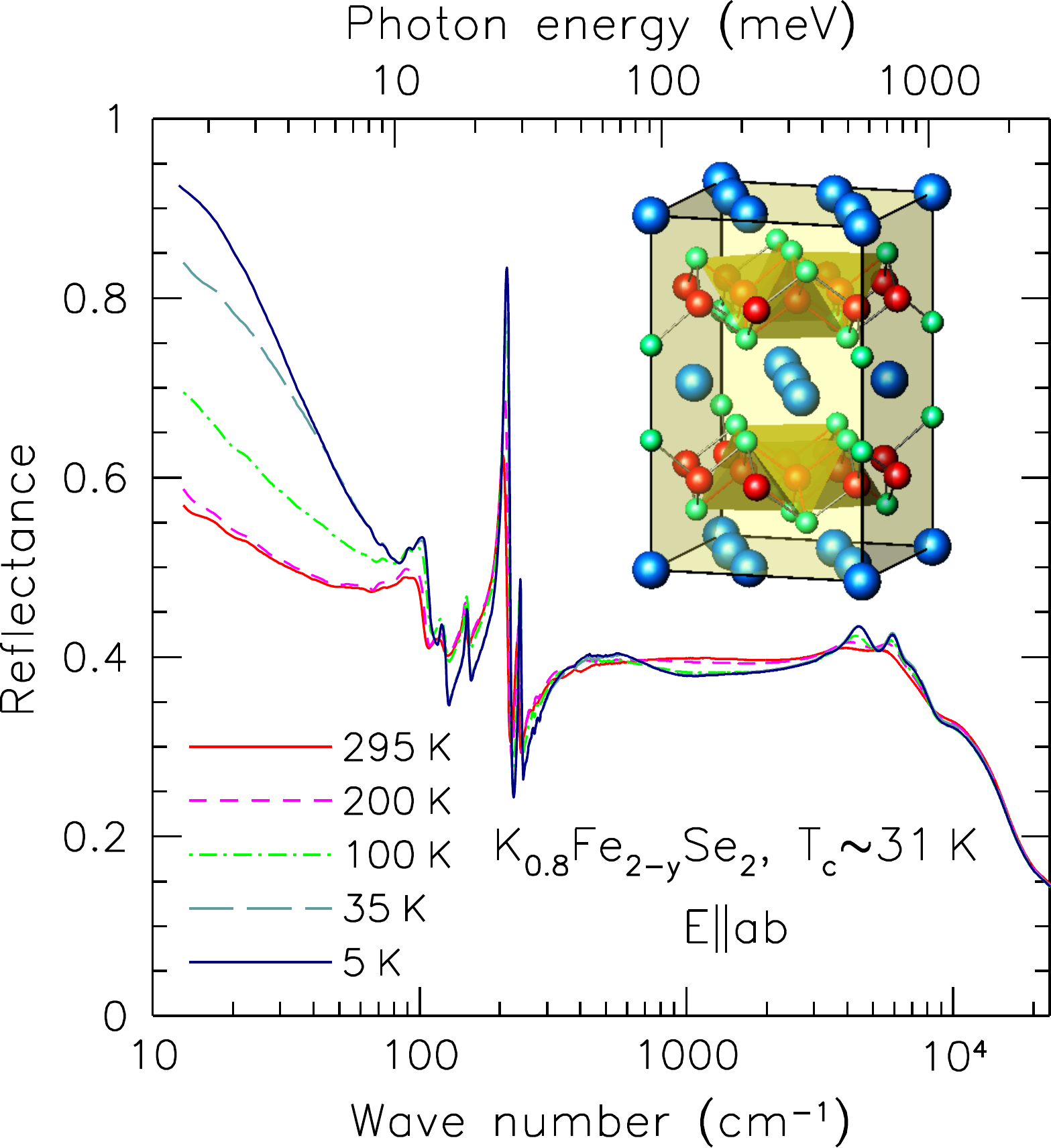}
\caption{The absolute reflectance over a wide frequency range for a cleaved
single crystal of K$_{0.8}$Fe$_{2-y}$Se$_2$ for light polarized in the
{\em a-b} planes at several temperatures above and below $T_c$.
Inset:  The unit cell in the $I4/m$ space group.}
\label{fig:reflec}
\end{figure}

%
% Experiment...
%
Large single crystals of K$_{0.8}$Fe$_{2-y}$Se$_2$ were grown using a vertical
unidirectional solidification method.  Pieces of potassium were added to precursor
FeSe to form the nominal composition and then placed into an alumina crucible and
sealed in a quartz tube; this tube was then resealed in another quartz
tube.  This arrangement was heated to $1050^\circ$C and then cooled to $700^\circ$C,
at which point the furnace was turned off and allowed to cool slowly to room temperature,
yielding mm-sized single crystals.  The room temperature resistivity of
$\sim 430$~${\rm m}\Omega\,$cm increases slightly upon cooling before beginning
to decrease below about 250~K, reaching a value of about 18~${\rm m}\Omega\,$cm
just before the sample becomes superconducting at $T_c = 31$~K. The value for
$T_c$ has been determined from magnetic susceptibility.
%
%
% Figure 2
%
\begin{figure}[t]
%\centerline{\includegraphics[width=2.8in]{figure2.eps}}%
\includegraphics[width=0.90\columnwidth]{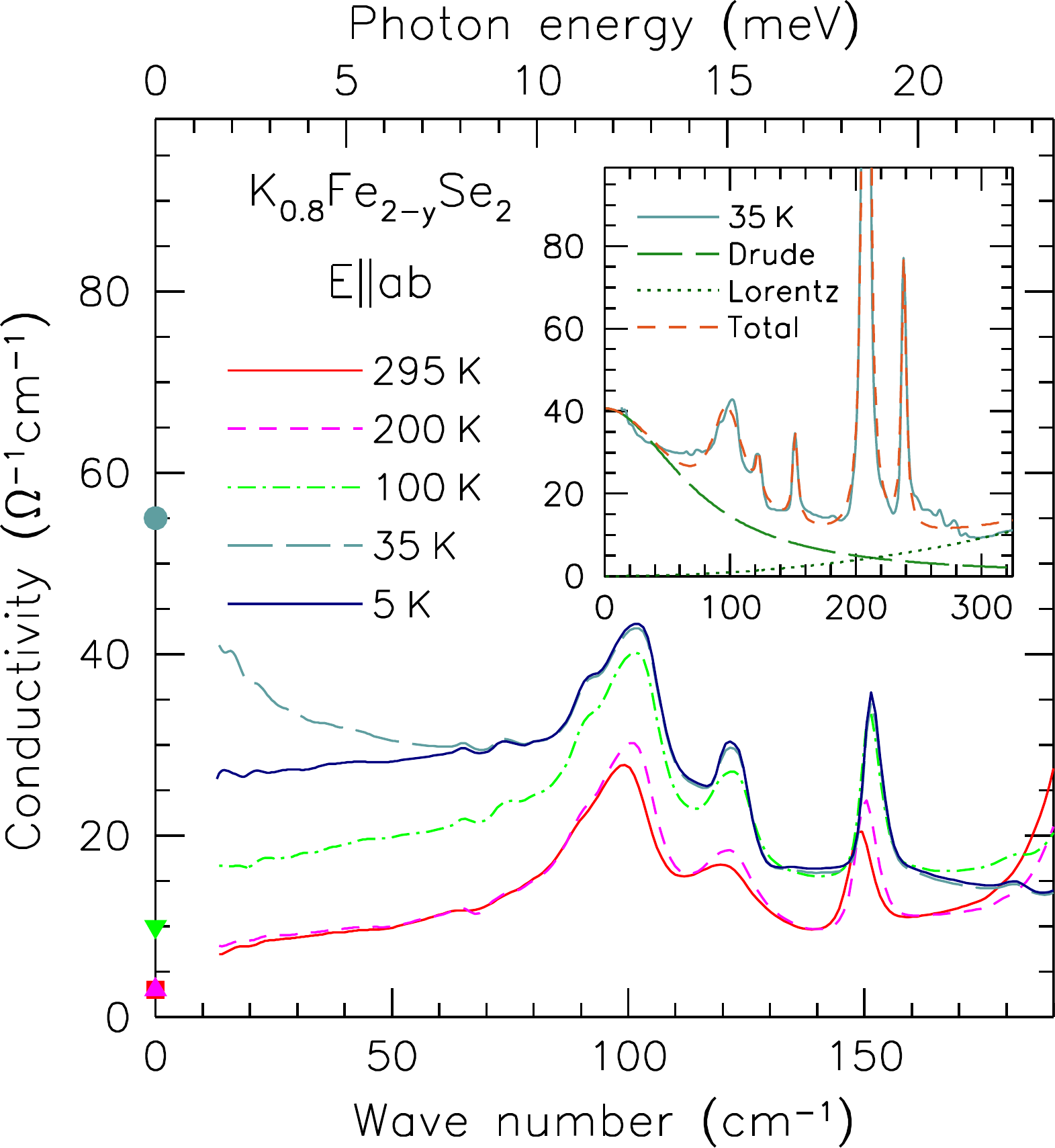}
\caption{The real part of the optical conductivity in the low-frequency
region for K$_{0.8}$Fe$_{2-y}$Se$_2$ for light polarized in the
{\em a-b} planes for several temperatures above and below
$T_c = 31$~K.  The polygons at the origin denote the values for the
conductivity determined from transport measurements on a crystal
from the same batch.
Inset: Drude-Lorentz fit to the conductivity at 35~K.}
\label{fig:sigma1}
\end{figure}
%
% Optics
%
A single crystal of K$_{0.8}$Fe$_{2-y}$Se$_2$ approximately $2\,{\rm mm}\times 2\,{\rm mm}
\times 100\,\mu{\rm m}$ was mounted on an optically-black cone and cleaved, revealing an
mirror-like surface; the sample was immediately transferred to a cryostat and placed under
vacuum.  The reflectance for light polarized in the iron-selenide ({\em a-b}) planes has
been measured using an {\em in-situ} overcoating technique\cite{homes93} from the terahertz
(1.5~meV) to the ultraviolet ($\sim 4$~eV) region for a variety of temperatures above and
below $T_c$, shown in Fig.~\ref{fig:reflec}.  The relatively low value for the reflectance
is characteristic of a poor metal and it is dominated by the infrared-active vibrations,
as well as other electronic features at higher energy.  The low-frequency reflectance
initially displays little temperature dependence, but begins to increase rapidly
below $\sim 200$~K, with an abrupt increase below $T_c$.  The complicated electronic
behavior contained in the reflectance may be seen more clearly in the optical
conductivity, which is determined from a Kramers-Kronig analysis of the
reflectance.\cite{dressel-book}

%
% Optical conductivity
%
The calculated conductivity is shown in the low-frequency region in Fig.~\ref{fig:sigma1}.
At room temperature the conductivity is dominated by a series of sharp features
associated with the infrared-active lattice modes, superimposed on a flat, incoherent
electronic background.   As the temperature is reduced, the vibrational features
narrow and increase slightly in frequency.  Below $\simeq 200$~K the low
frequency conductivity begins to gradually increase, until just above $T_c$ the
response may be described reasonably well by using a Drude-Lorentz model
for the complex dielectric function
\begin{equation}
  \tilde\epsilon(\omega) = \epsilon_\infty - {{\omega_{p,D}^2}\over{\omega^2+i\omega/\tau_D}}
    + \sum_j {{\Omega_j^2}\over{\omega_j^2 - \omega^2 - i\omega\gamma_j}},
\end{equation}
where $\epsilon_\infty$ is the real part of the dielectric function at high
frequency, $\omega_{p,D}^2 = 4\pi ne^2/m^\ast$ and $1/\tau_D$ are the plasma frequency
and scattering rate for the delocalized (Drude) carriers, respectively, and $m^\ast$
is an effective mass tensor; $\omega_j$,
$\gamma_j$ and $\Omega_j$ are the position, width, and oscillator strength of
the $j$th vibration.  The complex conductivity is $\tilde\sigma(\omega) =
\sigma_1 +i\sigma_2 = -i\omega [\tilde\epsilon(\omega) - \epsilon_\infty ]/4\pi$.
%
% Fit and high-frequency features...
%
The inset in Fig.~\ref{fig:sigma1} shows the result of a non-linear least-squares
fit to the conductivity at 35~K using a single Drude component as well as a number
of Lorentz oscillators to reproduce the narrow phonon features and the broader
excitations observed at high frequency.
%
% Table I - ab plane parameter fits
%
\begin{table}[t]
\caption{The fitted vibrational parameters for the in-plane infrared active
lattice modes in the optical conductivity of K$_{0.8}$Fe$_{2-y}$Se$_2$ at
35~K, where $\omega_j$, $\gamma_j$ and $\Omega_j$ are the frequency, width
and oscillator strength of the $j$th mode.  The errors are estimated from the
covariance and are indicated in parenthesis.  All units are in cm$^{-1}$.}
\begin{ruledtabular}
\begin{tabular}{c c c}
 \ \ $\omega_j$ ($\delta\omega_j$) \ &
 \ \ $\gamma_j$ ($\delta\gamma_j$) \ &
 \ \ $\Omega_j$ ($\delta\Omega_j$) \ \\
  \hline
%
% Ellab
%
% spacing controls
%
 \ \  65.2 (0.2) &   2.5 (0.3)  &   10  (2) \ \ \\
 \ \  73.6  (0.2) &  3.7  (0.4) &  15   (3) \ \ \\
 \ \  93.7  (0.2) & 18.1  (3.4) & 118  (17) \ \ \\
 \ \ 102.3  (0.6) & 10.5  (0.9) & 103  (17) \ \ \\
 \ \ 121.9  (0.6) &  9.2  (0.8) &  78   (6) \ \ \\
 \ \ 151.7  (0.3) &  3.8  (0.5) &  67   (6) \ \ \\
 \ \ 181.9  (0.3) &  4.5  (0.2) &  16   (2) \ \ \\
 \ \ 208.3  (0.1) &  4.4  (0.1) & 322   (8) \ \ \\
 \ \ 238.3  (0.1) &  4.3  (0.2) & 131   (6) \ \ \\
 \ \ 267.1  (0.2) &  2.9  (0.2) &  22   (3) \ \ \\
 \ \ 278.6  (0.2) &  5.2  (0.2) &  32   (3) \ \ \\
\end{tabular}
\end{ruledtabular}
%\footnotetext[1] { }
\label{tab:fit}
\end{table}

%
% Group theory & vibrations
%
The large number of vibrations observed in this material (listed in
Table~\ref{tab:fit}) is surprising given that in tetragonal BaFe$_2$As$_2$
($I4/mmm$) only two in-plane infrared active $E_u$ modes are expected and
observed, and that even below the magnetic and structural transition the
orthorhombic distortion ($Fmmm$) only yields an additional two modes,\cite{akrap09}
far fewer than observed here.  However, it has been remarked
that the $I4/m$ tetragonal unit cell for K$_{0.8}$Fe$_{2-y}$Se$_2$ (shown in
the inset of Fig.~\ref{fig:reflec}) is larger and more complicated than the
related iron-arsenic structure.\cite{bao11,bacsa11,zavalij11}  We have
determined the the irreducible vibrational representation using the
$I4/m$ space group,
$$
  \Gamma_{vib} = 9A_{1g} + 9B_{1g} + 9E_{g} + 7A_{u} + 7B_{u} + 10E_{u}.
$$
The $g$ modes are Raman active and the $B_u$ modes are silent; only the
$A_{u}$ and $E_u$ vibrations are infrared active along the {\em c} axis
and in the {\em a-b} planes, respectively.  The large number of infrared-active
modes is in agreement with observation; however, while only 10 modes are
predicted, a minimum of 11 modes are observed (two additional features,
weak shoulders associated with the two strong modes at $\simeq 208$ and
238~cm$^{-1}$ are difficult to fit and have not been included).  This
suggests that some of the observed modes may be due to a secondary
phase, the activation of Raman modes due to disorder, or both.

\begin{figure}[tb]
%
% manuscript
%
%\centerline{\includegraphics[width=2.8in]{figure3.eps}}%
\includegraphics[width=0.90\columnwidth]{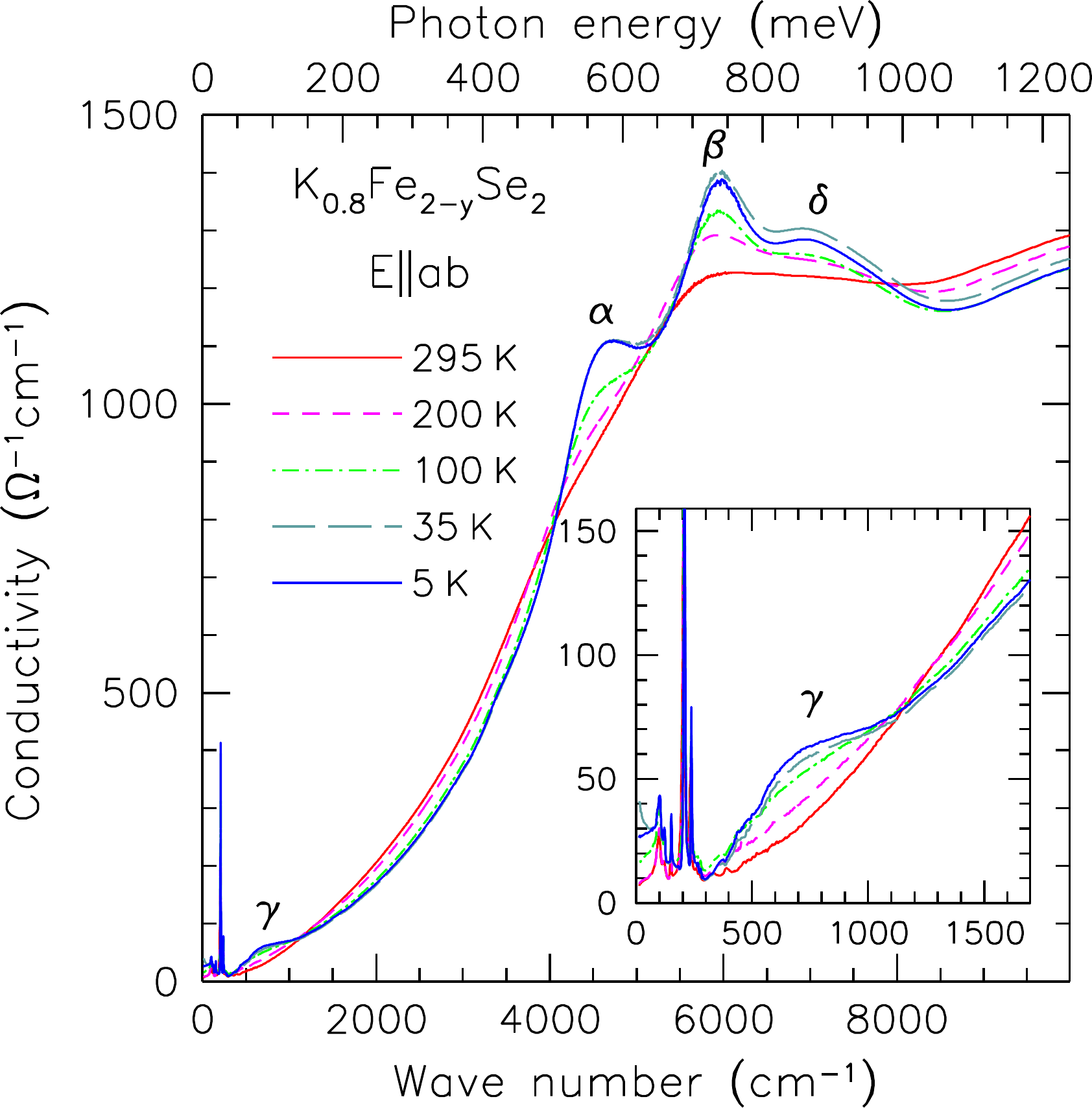}
\caption{The real part of the optical conductivity for light polarized in
  the {\em a-b} planes of K$_{0.8}$Fe$_{2-y}$Se$_2$ over a wide frequency range
  for several temperatures above and below $T_c$.  Two previously observed
  features are labeled $\alpha$ and $\beta$; two new features at low and
  high frequency are denoted as $\gamma$ and $\delta$, respectively.
  Inset: The detailed behavior of the $\gamma$ feature.}%
\label{fig:sigma2}
\end{figure}

%
% Structure at high frequency
%
In addition to the lattice modes, there are a number of broad features observed at
high frequency, shown in Fig.~\ref{fig:sigma2}.  The prominent features at
$\simeq 4600, 5900$ and 7100~cm$^{-1}$ have been observed in optical
studies of the K and Rb doped iron selenides\cite{chen11,charnukha12} and
are labeled $\alpha$, $\beta$ and $\delta$, respectively.   We also observe an
unusual low-energy feature at $\simeq 760$~cm$^{-1}$, denoted $\gamma$.
The three high-frequency features are thought to be related to spin-controlled
interband transitions.\cite{chen11,charnukha12}  Given that the 760~cm$^{-1}$
feature displays the same temperature dependence, it is possible that
it may have a similar origin.  The spectral weight associated with these features
increases at low temperature; however, this is compensated by a decrease in adjacent
regions so that the total spectral weight $N(\omega_c) = \int_{0^+}^{\omega_c}
\sigma_1(\omega)\,d\omega$ is constant when $\omega_c$ is sufficiently large
($\omega_c \gtrsim 1.5$~eV).

%
% Figure 3
%
%

%
% Discussion of normal and superconducting states...
%
The Drude parameters $\omega_{p,D} \simeq 430\pm 20$~cm$^{-1}$ and $1/\tau_D
\simeq 70\pm 5$~cm$^{-1}$ determined from the fit to the optical conductivity at
35~K \cite{note1} reveal that the Drude plasma frequency is more than an order of
magnitude smaller than what is observed in comparable iron-pnictide \cite{tu10}
or iron-chalcogenide materials.\cite{homes10}  A generalized-Drude
model\cite{allen77,puchkov96} indicates that the enhancement of the effective
mass at 35~K is rather small, $m^\ast(\omega\rightarrow 0)/m_e \simeq 2$.
This corresponds to a dilute carrier concentration of $n \simeq
4 \times 10^{18}$~cm$^{-3}$, considerably less than the estimates based on
NMR measurements.\cite{torchetti11}  Given the small value for $n$, it is
reasonable to ask whether or not this material is homogeneous. Just above
$T_c$ we note that $\sigma_{dc} = \sigma_1(\omega\rightarrow 0) \simeq
44$~$\Omega^{-1}$cm$^{-1}$, or $\rho_{dc} \simeq 23$~m$\Omega\,$cm.
The highly-anisotropic nature of these materials\cite{h-wang11} suggests
that the transport be described in terms of a sheet resistance $R_\Box
=\rho_{dc}/d$, where $d$ is the separation between the iron-chalcogenide
sheets.  Using $d\simeq 7$~\AA\cite{bao11,bacsa11,zavalij11} yields
$R_\Box \simeq 320$~k$\Omega$.  This value is significantly
larger than the threshold for the superconductor-insulator
transition\cite{strongin70} observed to occur close to $R_\Box = h/4e^2 \simeq
6.9$~k$\Omega$, suggesting that this material may not be homogeneous.

\begin{figure}[t]
%\centerline{\includegraphics[width=2.8in]{figure4.eps}}
\includegraphics[width=0.90\columnwidth]{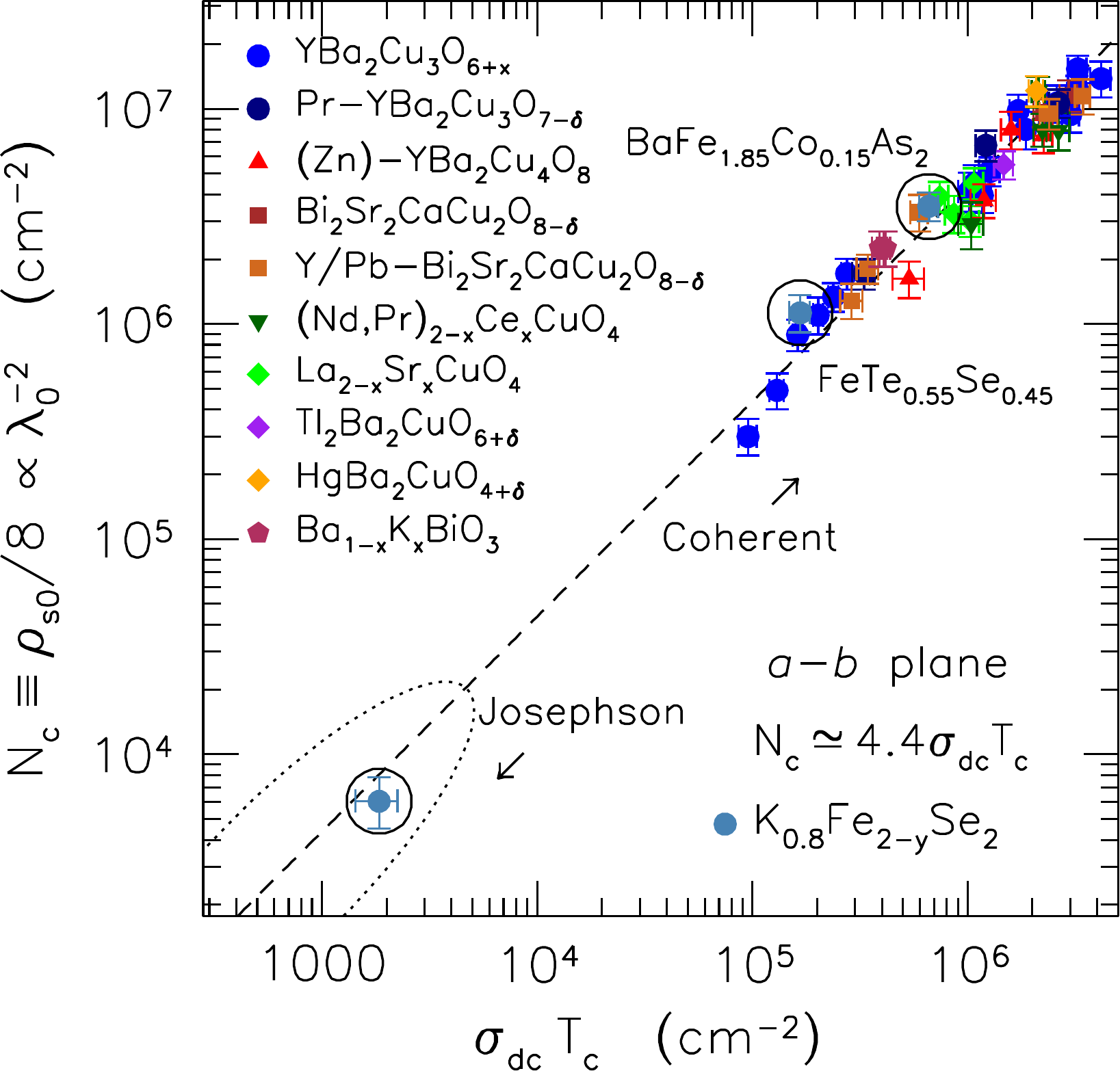}
\caption{The log-log plot of the spectral weight of the superfluid
 density $N_c \equiv \rho_{s0}/8$ vs $\sigma_{dc}\,T_c$ in the
 {\em a-b} planes for a variety of electron and hole-doped cuprates
 compared with BaFe$_{1.85}$Co$_{0.15}$As$_2$ \cite{tu10},
 FeTe$_{0.55}$Se$_{0.45}$ \cite{homes10}, and K$_{0.8}$Fe$_{2-y}$Se$_2$
 (this work), highlighted by the circles.  The dashed line corresponds
 to the general result for the cuprates $\rho_{s0}/8 \simeq 4.4
 \sigma_{dc} T_c$, while the dotted line denotes the region
 of the scaling relation typically associated with Josephson
 coupling along the {\em c} axis.}
\label{fig:scale}
\end{figure}

%
% SC response
%
Below $T_c$ in the superconducting state the conductivity
shows a characteristic suppression below $\simeq 65$~cm$^{-1}$.
The strength of the condensate may be determined from the
Ferrell-Glover-Tinkham sum rule, $N(\omega_c, T\gtrsim T_c) -
N(\omega_c, T\ll{T_c}) = \omega_{p,S}^2/8$, where $\omega_{p,S}$ is the
superconducting plasma frequency and $\omega_c$ is a cut-off frequency, allowing
the superfluid density $\rho_{s0}\equiv\omega_{p,S}^2$ to be calculated.
The small value for the missing area, the large amount of residual conductivity,
and the proximity to a number of phonon modes can all be significant sources
of uncertainty for this sum rule.  However, these difficulties can be
overcome by using the alternative method $\omega_{p,S}^2 = \omega\,\sigma_2(\omega)$
in the $T\ll T_c$, $\omega \rightarrow 0$ limit.\cite{jiang96}
The value $\omega_{p,S} \simeq 220\pm 20$~cm$^{-1}$ (yielding an
effective penetration depth $\lambda_{eff} \simeq 7.2\pm 0.7\,\mu$m) is
an order of magnitude smaller than what is observed in other
iron-arsenic\cite{tu10} and iron-chalcogenide\cite{homes10} superconductors.
It has been pointed out that a number of the iron-based superconductors fall
on the scaling relation initially observed for the cuprate superconductors,
$\rho_{s0}/8 \simeq 4.4\,\sigma_{dc}\,T_c$,\cite{homes04} shown in
Fig.~\ref{fig:scale}.
%
% Figure 4
%
%\begin{figure}[htbp]
The inhomogeneous nature of this material makes it difficult to associate
a given value for $\sigma_{dc}$ with the superconducting regions.  If we
consider the residual conductivity to originate mainly from a weakly-metallic
component that does not contribute to superconductivity, then we can
estimate $\sigma_{dc} \approx \sigma_1(T\gtrsim T_c) -\sigma_1(T\ll T_c)
\simeq 18$~$\Omega^{-1}$cm$^{-1}$ in the $\omega \rightarrow 0$ limit; we
attach the provision that there is a considerable amount of uncertainty
associated with this estimate.  Using these values, K$_{0.8}$Fe$_{2-y}$Se$_2$
does indeed fall on the scaling line in Fig.~\ref{fig:scale}; however, it
does so in a region associated with the response along the {\em c} axis
in the cuprates where the superconductivity is due to Josephson coupling
between the copper-oxygen planes.  This is further evidence that the
material is inhomogeneous and that it constitutes a Josephson phase,
in agreement with other recent results.\cite{yuan11,charnukha12,
z-wang11,ricci11a,ricci11b,shermadini11}
While it is tempting to associate the decrease in the low-frequency
conductivity below $T_c$ with the formation of a superconducting energy gap
$2\Delta \simeq 8$~meV, this value is significantly smaller than the ARPES
estimates of $2\Delta \simeq 16$~meV.\cite{x-wang11,zhang11,qian11}
Moreover, the previous statement that the superconductivity in this material
is due to Josephson coupling implies that the energy scale for changes in the
reflectance and conductivity should occur in the region of the renormalized
superconducting plasma frequency, $\tilde\omega_{p,S} = \omega_{p,S}/
\sqrt{\epsilon_{\rm FIR}}$; allowing that $\omega_{p,S} \simeq 220$~cm$^{-1}$
represents an average value for a distribution of frequencies \cite{yuan11} and
given the experimentally-determined value $\epsilon_{\rm FIR} \simeq 18$ at 50~meV,
then $\tilde\omega_{p,S} \simeq 52$~cm$^{-1}$ is reasonably close to the changes
observed in the conductivity below about 65~cm$^{-1}$.

%
% Summmarize this work...
%
In summary, the in-plane optical conductivity of K$_{0.8}$Fe$_{2-y}$Se$_2$
($T_c = 31$~K) is incoherent at room temperature where it is dominated by
infrared-active lattice modes and other high-frequency bound excitations.
Just above $T_c$ a coherent, Drude-like, response emerges, but the fitted
value $\omega_{p,D} \simeq 430$~cm$^{-1}$ is quite small.  Moreover,
$R_\Box \simeq 320$~k$\Omega$ at 35~K, well above the threshold for the
superconductor-insulator transition observed at $R_\Box = h/4e^2 \simeq
6.9$~k$\Omega$.  Below $T_c$ the superconducting plasma frequency $\omega_{p,S}
\simeq 220$~cm$^{-1}$ is more than an order of magnitude smaller than what
is observed in other iron-based superconductors.  This material falls on the
scaling line $\rho_{s0}/8 \simeq 4.4\,\sigma_{dc}\, T_c$, but does so in a
region associated with Josephson coupling along the poorly-conducting {\em c}
axis in the cuprate superconductors.  Taken together, the normal and
superconducting state properties suggest an inhomogeneous, phase separated
material, in which the superconductivity is due to Josephson coupling.

%
% Acknowledgements...
%
We are grateful to C. Petrovic for suggesting this series of experiments.
We would like to acknowledge useful discussions with W. Bao, A. V. Chubukov,
Q. Li, M. Rechner, A. M. Tsvelik, and N. L. Wang.
Research supported by the U.S. Department of Energy, Office of
Basic Energy Sciences, Division of Materials Sciences and Engineering
under Contract No. DE-AC02-98CH10886.  Z.~X. and J.~W. are supported
by the Center for Emergent Superconductivity, an Energy
Frontier Research Consortium supported by the Office of
Basic Energy Science of the Department of Energy
%

%
%%%%%%%%%%%%%%%%%%%%%%%%%%%%%%%%%%%%%%%%%%%%%%%%%%%%%%%%%%%%%%%%%%%%%%%%%%%%%%
%
% References
%
%\bibliography{kfe2se2}

\begin{thebibliography}{36}
\bibitem [{not()}]{note1}%
  \BibitemOpen
  \href@noop {} {}\bibinfo {note} {A further mode at $\simeq 11,000$~cm$^{-1}$
  has been included in the fits.}\BibitemShut {Stop}%
\end{thebibliography}
%
%merlin.mbs apsrev4-1.bst 2010-07-25 4.21a (PWD, AO, DPC) hacked
%Control: key (0)
%Control: author (8) initials jnrlst
%Control: editor formatted (1) identically to author
%Control: production of article title (-1) disabled
%Control: page (0) single
%Control: year (1) truncated
%Control: production of eprint (0) enabled
%

\end{document}